\documentclass[]{mn2e}
\usepackage{graphics}
\usepackage{epsf}
\usepackage{subfigure}

\begin{document}

% CHANGE LATER
%\thesaurus{04(11.04.1; 08.22.1)}

\title{Period Color and Amplitude Color relations for MACHO project LMC RR Lyraes}
\author[Kanbur and Fernando]{S. M. Kanbur$^{1}$\thanks{Email: shashi@astro.umass.edu} and I. Fernando$^{1}$ 
\\
$^{1}$Department of Astronomy, University of Massachusetts
\\
Amherst, MA 01003, USA}

%\offprints{S. Kanbur \\
%email: {\tt shashi@astro.umass.edu}}

\date{Received XX XXX 2004 / Accepted XX XXX 2004}

%\titlerunning{RR Lyraes, LMC}
%\authorrunning{Kanbur \& Fernando}
\maketitle

\begin{abstract}
In this paper, we analyze period color and amplitude color relations at
minimum, mean and maximum $V$ band light
for 6391 RRab stars in the Large Magellanic Cloud obtained by the
MACHO project. Specifically, we find that
color and amplitude are nearly independent of period at minimum light
but that there exists a definite relation between period and color and
amplitude and color at maximum light. These two
properties are easily explained by the application of the
Stefan Boltzmann law and the interaction of
the photosphere and hydrogen ionization front at minimum light.
When we examine the slope
of the period color relation as a function of phase, we find that the slope
varies significantly with phase and
is small for a wide range of phases around minimum light. This suggests that
another factor that needs to be considered when trying to understand
RR Lyrae observed properties is their behavior at different phases
during a pulsation cycle.

\end{abstract}


%***********************************************************
%   SECTION 1: INTRODUCTION
%***********************************************************

\section{Introduction}

Kanbur and Ngeow (2004, hereafter KN) and Kanbur, Ngeow and Buchler (2004, hereafter KNB) and Ngeow and Kanbur (2004b) analyzed Cepheid light curves in the Galaxy, LMC and SMC to
derive period color (PC) and amplitude color relations (AC) as a function of phase. They analyzed their results in the
context of the work of Simon Kanbur and Mihalas (1993, hereafter SKM). SKM presented the following equation, derived from an application of the Stefan
Boltzmann law at maximum and minimum light assuming that radius fluctuations
are small,
$$ \log L_{max} - \log L_{min} = 4(\log T_{max} - \log T_{min}) + const.,\eqno(1)$$
where $L_{max},L_{min}, T_{max}, T_{min}$ are photospheric
luminosities and temperatures at these phases.
If we consider wavelength bands which are good indicators of
temperature, say $X,Y$, so that $\log T = a + b(X-Y)$, then equation (1) becomes
$$X_{min} - X_{max} = -10.b((X-Y)_{max} - (X-Y)_{min}) + const., \eqno(2)$$
where $(X-Y)_z = X_z-Y_z$, and $Y_z$ denotes the value of $Y$ at $X_z$. If, for some physical reason, there is a
flat relation between period and either $T_{max}$ or $T_{min}$, then
equations (1) and (2) predict that there will be a relation between
amplitude and the $T_{min}$ or $T_{max}$ respectively. Assuming our
observations are in bands which are good indicators of temperature, then
a flat relation between period and color at max/min is related to a
relation between amplitude and color at min/max light.

Whilst Cepheids display a flat relation between period and color
at maximum light (KN, KNB),
Sturch (1967) and Clementini et al (1995) published evidence that fundamental
mode RR Lyrae stars (RRab) are such that they follow a flat relation between
period and effective temperature at {\it minimum} light. Kanbur (1995)
and Kanbur and Phillips (1996) again used radiative
hydrodynamical models of
RRab stars to provide credible evidence that the physical reason for this
was again the interaction of the HIF and photosphere. For RRab stars,
this interaction occurs at minimum light. This difference, between Cepheids and
RR Lyraes, in the phase
at which an interaction between the photosphere and HIF occurs is caused by the
hotter temperatures and lower $L/M$ ratios relevant for RRab stars (Kanbur 1995).
However, these
previous studies did not look for evidence of amplitude-color (AC)
relations.
Extensive data in two wavebands for 6391 RR Lyrae
stars in the LMC discovered by the MACHO project (Alcock et al, 2003) provides an ideal
opportunity both to verify earlier observational results and look for AC relations
in RRab stars. This is the main thrust of this paper.

\section{The Data}
The data considered were obtained by the MACHO project and initially
analyzed in Alcock et al (2003). These data consist of,
on average, some 400-800 points per star in both MACHO $V$ and $R$
bands and were kindly supplied by Kovacs (2004). There are 6391
stars considered in our sample.

We used the periods published in Alcock et al (2003). In a future paper, we will
compare these periods with those calculated by the  Lomb-Scargle method (Lomb 1976, Scargle 1982). 
We then used this published period
to fit a sixth order Fourier series to both $V$ and $R$ band data.
Figure 1 shows a representative star from our sample. The solid line
displays the Fourier fit and the open squares the original $V$ band data.
We estimated quantities at maximum, mean, minimum $V$ band light and colors using the
Fourier fits to both $V$ and $R$ band data.
We removed stars whose periods did not lie between 0.3 and 0.9 days
and whose $V-R$ color did not lie between 0 and 0.6. We also
removed stars not classified as fundamental mode RRab stars by Alcock et al
(2003). This left 4830 stars
which we studied further. We leave the detailed treatment of the remaining stars
for another paper. 
We emphasize that we did {\it not} correct our $V-R$
colors for extinction.

\section{Results}

Our period color (PC) and amplitude-color relations
at maximum, mean and minimum $V$ band light are given in figures 2-4
and figures 5-6 respectively. The two rows of table 1 provide the
slopes, together with their standard errors, of straight line fits
to these data for PC and AC relations. We see that the PC relation at
minimum light has a slope close to zero ($0.041\pm0.012$) whilst the
PC relation at maximum light has a significantly different slope
($0.423\pm 0.015$). The AC relation also has a slope
close to zero at minimum light ($-0.0005\pm0.003$). This slope becomes
significantly different at maximum light ($-0.180\pm0.002$). In both
PC and AC cases, the behaviour at mean light is intermediate between the
behaviour at maximum/minimum light. The large number of stars means that the slopes
are quite accurately determined.
These results are robust to changes in the color cuts adopted in the previous section.
For example, the difference in the slopes of the PC and AC relations at maximum and minimum light
is still present even when we exclude stars whose $V-R$ colors do not lie between 0.2 and 0.4.
Our results are similarly  robust to changes in the adopted period cuts. 
Figures 2-7 and table 1 are completely consistent with the theory developed
in SKM, Kanbur (1995), Kanbur and Phillips (1996) and 
Kanbur, Ngeow and Buchler (2004).

Figure 8 provides a plot of the slope in a PC relation
against the phase at which the PC relation is evaluated, where we
specify that phase 0 is minimum $V$ band light. We see clearly
that there is a large range in slope and that this slope
decreases to its lowest value at minimum light and is
in fact smaller than 0.1 for a range of phases ($\approx 0.4$)
around minimum light. Further, the slope at mean light given in table 1
is very close to the figure obtained when taking the straight average of the
slope at each phase point in figure 8. Hence PC relations at mean light
{\it will} be affected by the behaviour of PC relations at other
phases and in particular by the physics of RRab stars at
minimum light. Figure 8 not only provides further evidence for our results
but also implies that in order to fully understand observed
properties of RRab stars, it is sometimes necessary to study those
properties as a function of phase.

Could not correcting for extinction be affecting our results?
Were this to be true, figures 2-7 imply that this omission would have
a rather drastic and systematic nature. For example in order to use this to "make"
the PC or AC relation at maximum light flat, would mean that extinction of the order of 0.1 - 0.2mags
affects about half of these stars. A proper accounting for extinction would probably make our
relations tighter.

Could an amplitude bias due to crowded field photometry affect our results?
Alcock et al (2004) have discussed how to estimate such a bias. They used
these estimates to correct estimates of the mean magnitude of first overtone
RRC
stars in the MACHO LMC sample. Their estimate of amplitude bias ranged from
-0.1
to -0.2 mags. Even if a large fraction of the stars in figure 5 had
amplitudes underestimated by 0.2mags, this would not be enough to remove the
non-zero slope in this figure.

We have produced similar PC/AC plots at these three phases for
all the other types of RR Lyrae stars identified by Alcock et al (2003).
BL1 and BL2 type Blazhko variables display similar behaviour. Results for the 36 RRc stars
in this sample are inconclusive. The stars classified as binary stars do not exhibit the
behaviour displayed in figures 3-7.

\section{Conclusions and Discussion}

By studying the large sample of RRab stars in the LMC provided by the
MACHO project, and using the MACHO $V-R$ colors, we have found compelling evidence that RRab stars in the
LMC display a PC and AC relation that is flat at minimum light but that has a significant slope at
maximum light. By considering
the PC and AC relations at minimum and maximum light in the context of SKM, KN and KNB, we also found
strong evidence that these observational features of RRab stars are caused by the
interaction of the photosphere and hydrogen ionization front though we caution that final
confirmation must await the construction of new hydrodynamic models of RRab stars in the LMC. In
particular we show convincing evidence of a relation between $V$
band amplitude
and $V-R$ color at maximum light such that higher amplitude stars are
driven to hotter temperatures at $V$ band maximum light. Galactic Cepheids
display the "opposite" property: a flat PC relation at maximum
light and an AC relation such that higher amplitude stars are driven to
cooler temperatures at minimum light (SKM, KN, KNB). Note that this is
more than just saying brightness fluctuations are caused predominantly by
temperature variations. This is indeed true for classical Cepheid variables, but in addition, given a way
to get a flat
PC relation at max/min light, then we will have an AC relation at the opposite
max/min phase. Further, since the PC relation for RRab stars
is flat for a large range of phases around minimum light,
this work suggests
that the study of the way the photosphere interacts with the
HIF as a function of phase and metallicity will provide deeper
insights into the observed properties of both RR Lyraes and Cepheids.
These projects are currently under way.

\begin{figure}
\hbox{\hspace{0.1cm}\epsfxsize=8.0cm \epsfbox{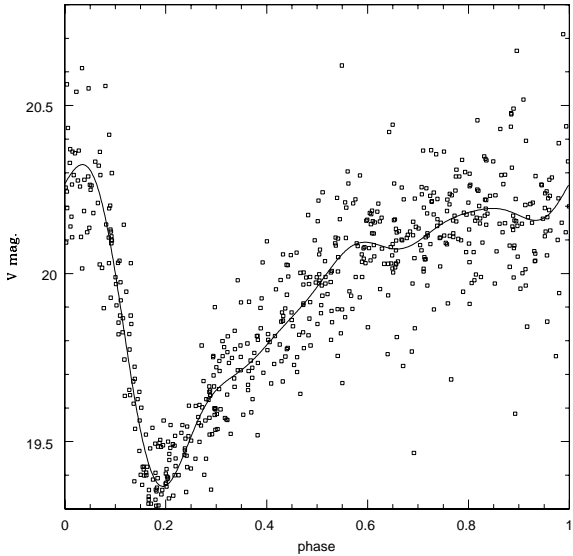}}
\caption{Fourier decomposition curve together with original V band data for a representative star in the sample.}
\label{ref}
\end{figure}

\begin{figure}
\hbox{\hspace{0.1cm}\epsfxsize=8.0cm \epsfbox{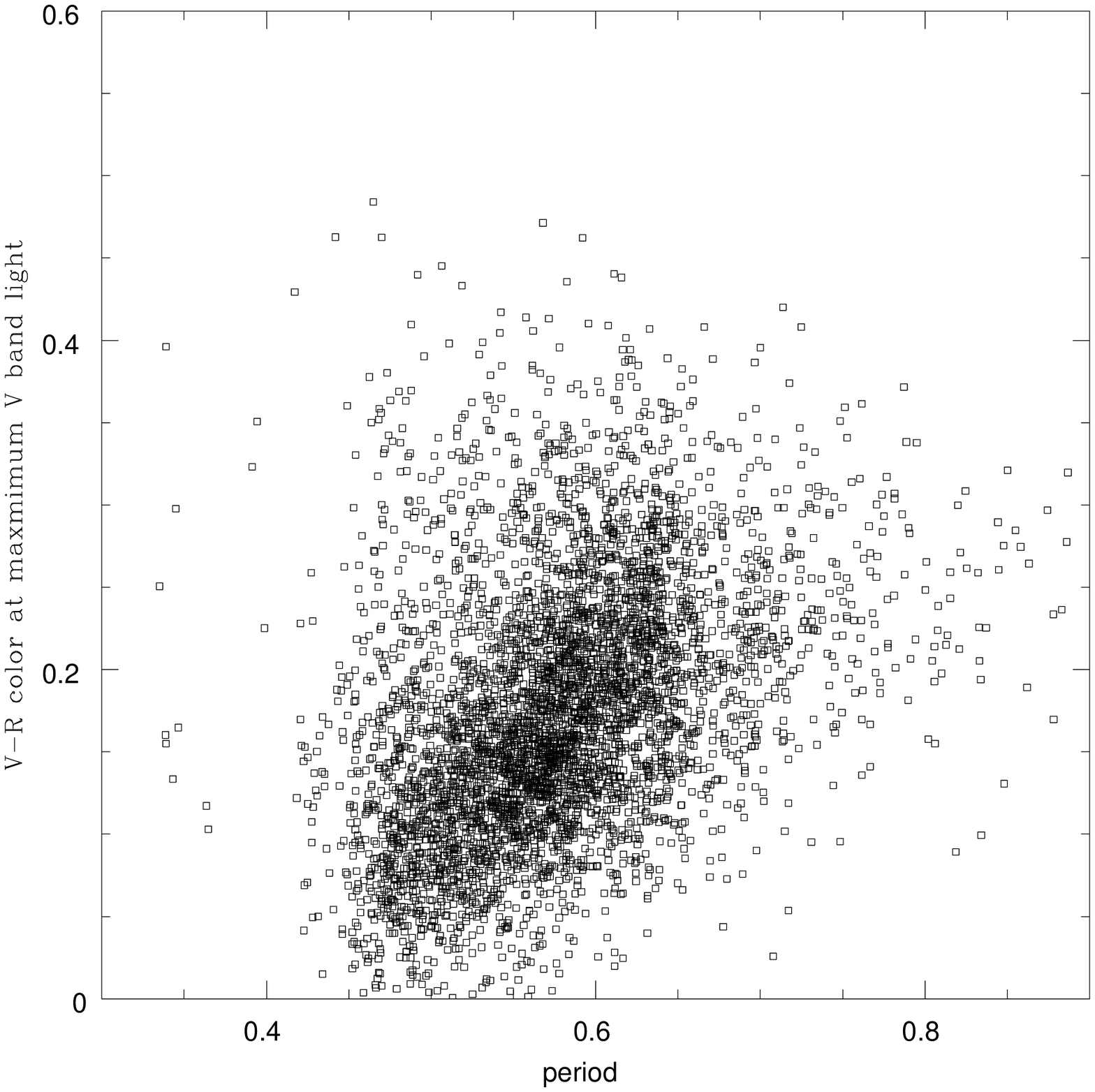}}
\caption{ Period against V-R color at maximum V band light.}
\label{ref}
\end{figure}

\begin{figure}
\hbox{\hspace{0.1cm}\epsfxsize=8.0cm \epsfbox{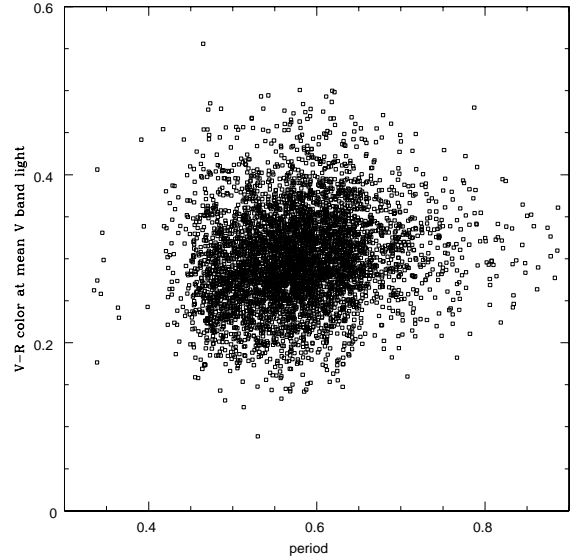}}
\caption{Period against V-R color at mean V band light.}
\label{ref}
\end{figure}

\begin{figure}
\hbox{\hspace{0.1cm}\epsfxsize=8.0cm \epsfbox{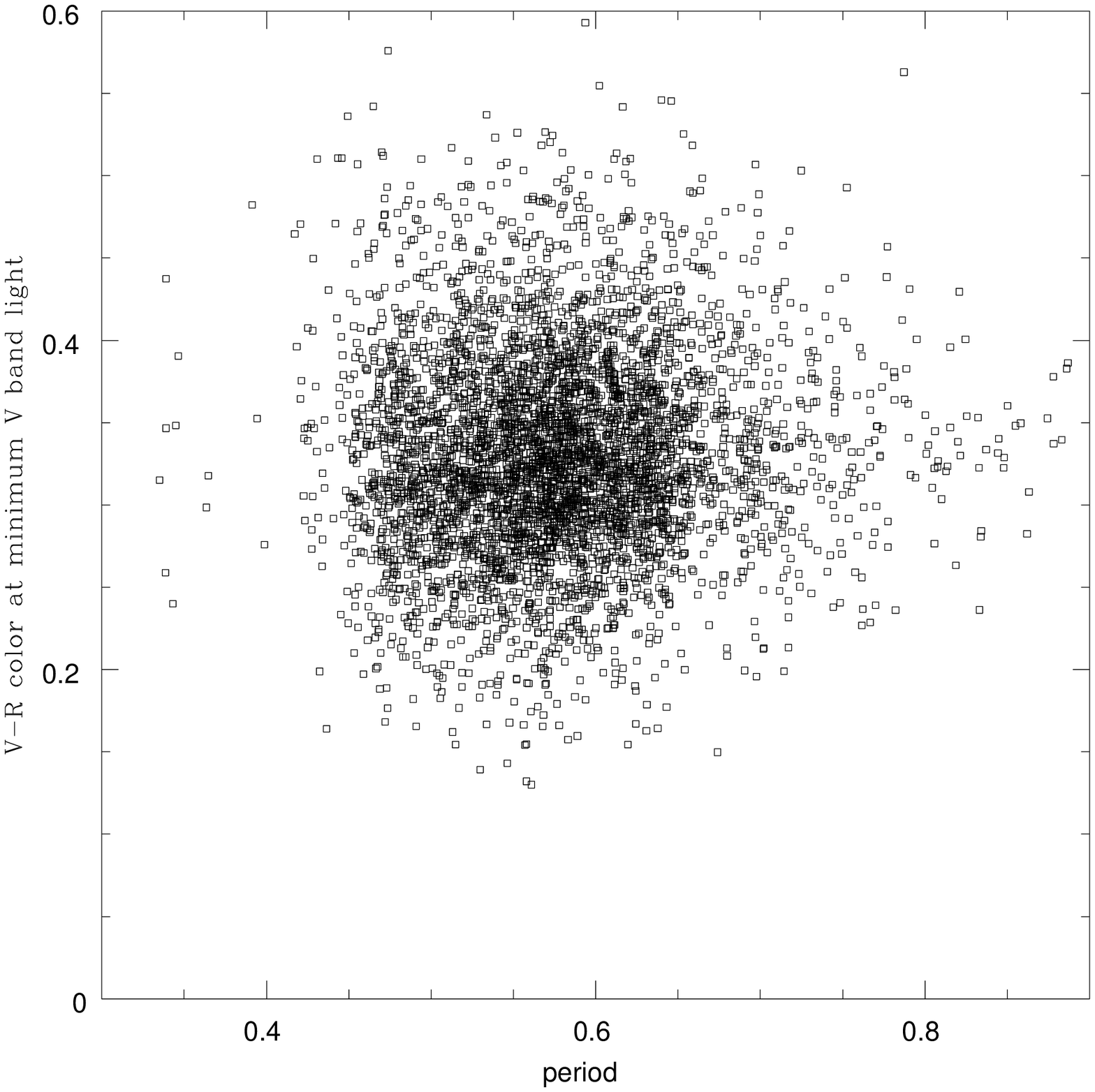}}
\caption{Period against V-R color at minimum V band light.}
\label{ref}
\end{figure}

\begin{figure}
\hbox{\hspace{0.1cm}\epsfxsize=8.0cm \epsfbox{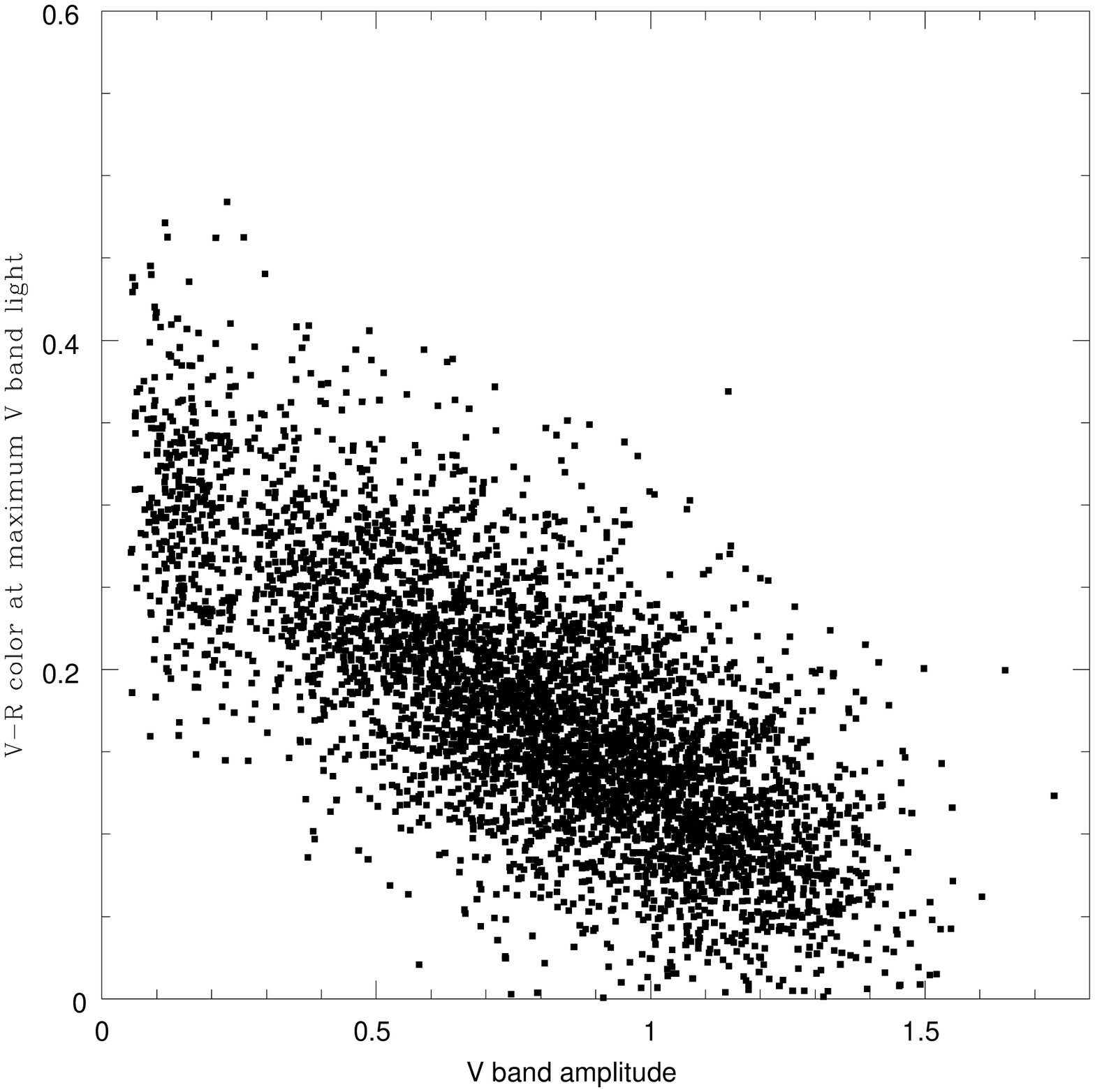}}
\caption{V band amplitude against V-R color at maximum V band light.}
\label{ref}
\end{figure}

\begin{figure}
\hbox{\hspace{0.1cm}\epsfxsize=8.0cm \epsfbox{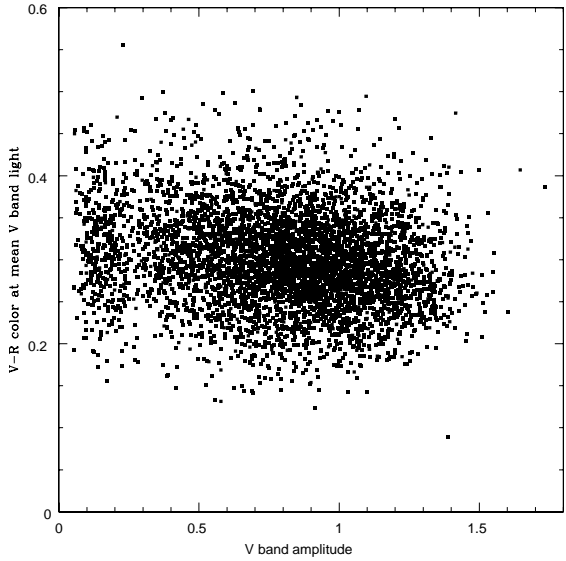}}
\caption{ V band amplitude against V-R color at mean V band light.}
\label{ref}
\end{figure}

\begin{figure}
\hbox{\hspace{0.1cm}\epsfxsize=8.0cm \epsfbox{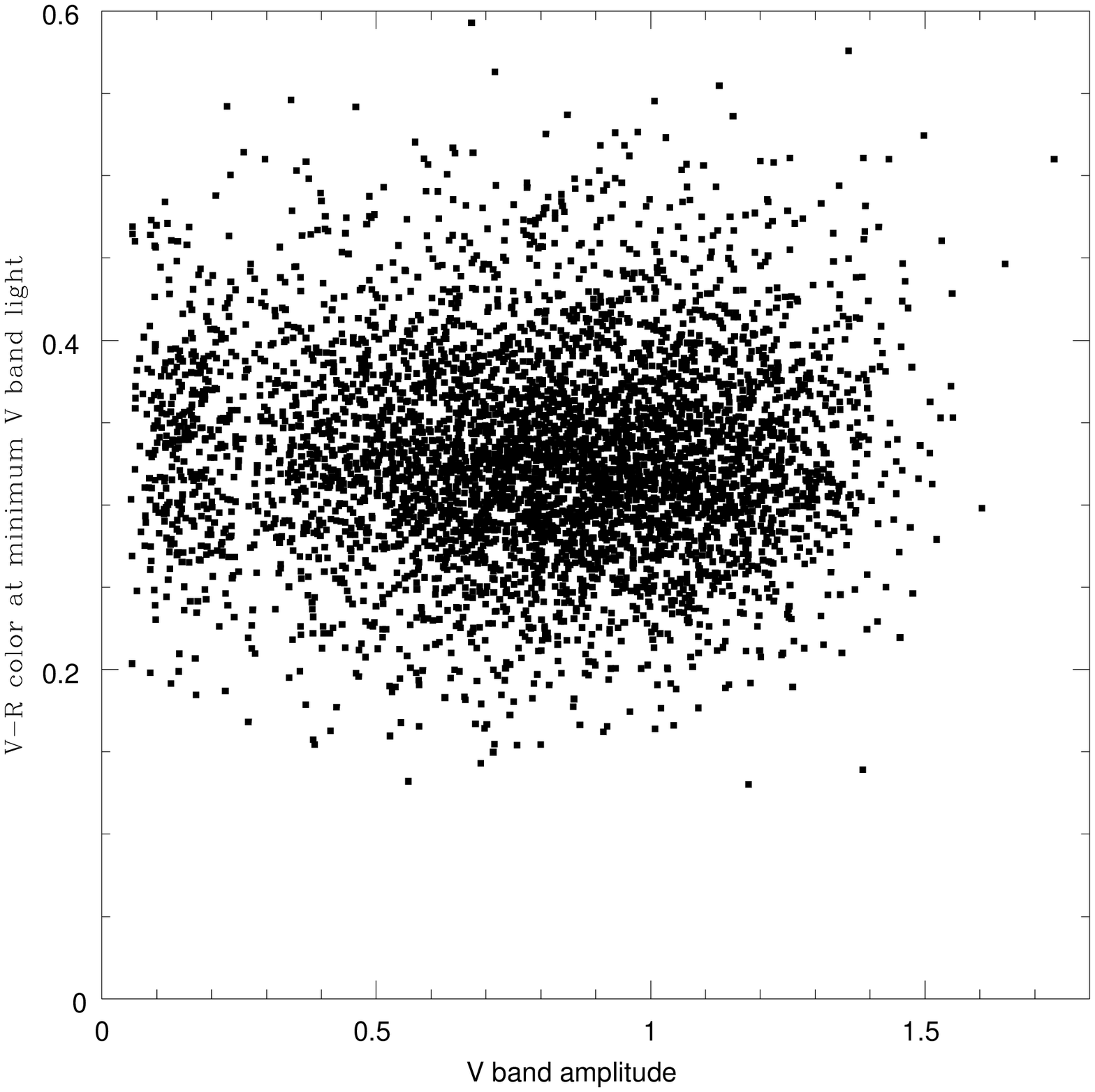}}
\caption{V band amplitude against V-R color at minimum V band light.}
\label{ref}
\end{figure}

\begin{figure}
\hbox{\hspace{0.1cm}\epsfxsize=8.0cm \epsfbox{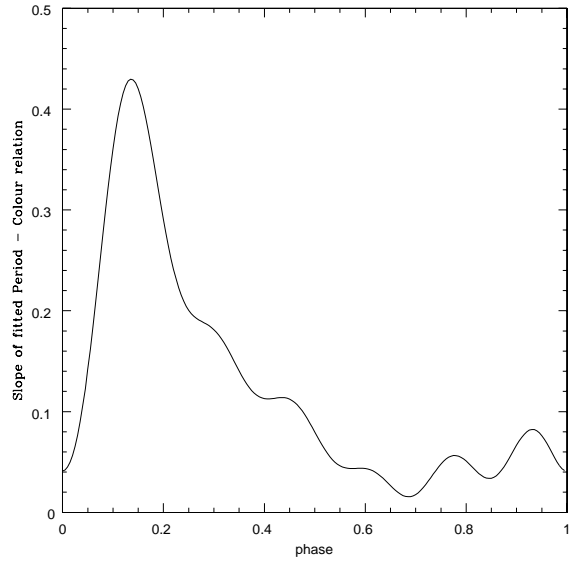}}
\caption{Plot of fitted phase against PC slope}
\label{ref}
\end{figure}

\begin{table}
\centering
\caption{ Period color and Amplitude color slope with error at maximum, mean and minimum V band light}
\label{tab1}
\begin{tabular}{lcccc} \\ \hline
&Minimum&Mean&Maximum \\ \hline
PC&$0.0411\pm0.012$&$0.118\pm0.011$&$0.423\pm0.015$ \\
AC&$-0.0005\pm0.003$&$-0.030\pm0.003$&$-0.180\pm0.002$\\
\hline
\end{tabular}
\end{table}

\section*{acknowledgments}

This paper utilizes public domain data obtained by the MACHO project,
jointly funded by the US Department of Energy through the University of
California, Lawrence Livermore National Laboratory under contract No. W-7405-Eng-48,
by the National Science Foundation through the Center for Particle Astrophysics of the
University of California under cooperative agreement AST-8809616, and by the
Mount Stromlo and Siding Spring Observatory, part of the Australian National University.

SMK thanks Geza Kovacs for providing the raw data and to D. Welch, K. Cook and D. Alves
for stimulating discussions. IF thanks the Massachusetts Space Grant Consortium for funding a
FCRAO summer internship in 2004 when this work was completed.

%**********************************************************
%   FIGURES
%**********************************************************

%\begin{figure*}
%\resizebox{12cm}{!}{\includegraphics{galcol.ps}}
%\hfill
%\parbox[b]{55mm}{
%\caption{Plots of log period against maximum (bottom), mean (middle) minimum (top) (V-I) color for the Galaxy.
%\label{fig1}} 
%}
%\end{figure*}

\end{document}